\documentclass[oneside]{ar2e}
\usepackage{bm}
\usepackage{amssymb}
\usepackage{fix-cm}
\usepackage{graphicx}
\usepackage{hyperref}

\usepackage{color}

\newcommand{\CP}{\ensuremath{CP}}
\newcommand{\tr}{\ensuremath{\mathop\mathrm{tr}}}
\newcommand{\Tr}{\ensuremath{\mathop\mathrm{Tr}}}
\newcommand{\alphas}{\ensuremath{\alpha_\mathrm{s}}}
\newcommand{\mup}{\ensuremath{m_\mathrm{u}}}
\newcommand{\mdn}{\ensuremath{m_\mathrm{d}}}
\newcommand{\mstr}{\ensuremath{m_\mathrm{s}}}
\newcommand{\mch}{\ensuremath{m_\mathrm{c}}}
\newcommand{\mbt}{\ensuremath{m_\mathrm{b}}}
\newcommand{\mtop}{\ensuremath{m_\mathrm{t}}}
\newcommand{\Mup}{\ensuremath{\bar{m}_\mathrm{u}}}
\newcommand{\Mdn}{\ensuremath{\bar{m}_\mathrm{d}}}
\newcommand{\Mstr}{\ensuremath{\bar{m}_\mathrm{s}}}
\newcommand{\Mch}{\ensuremath{\bar{m}_\mathrm{c}}}
\newcommand{\Mbt}{\ensuremath{\bar{m}_\mathrm{b}}}

\newcommand{\case}[2]{\ensuremath{{\textstyle\frac{#1}{#2}}}}
\newcommand{\half}{\ensuremath{{\textstyle\frac{1}{2}}}}

\usepackage{floatmyboat}

\begin{document}

\hyphenation{bot-tom-onium charm-onium quark-onium}

\jname{Annu. Rev. Nucl. Part. Sci.}
\jyear{2012}
\jvol{62}
\ARinfo{1056-8700/97/0610-00}

\title{Twenty-first Century Lattice Gauge Theory: \\ Results from the QCD Lagrangian}

\markboth{21st Century Lattice QCD}{21st Century Lattice QCD}

\author{Andreas S. Kronfeld
\affiliation{Theoretical Physics Department, Fermi National Accelerator Laboratory, 
P.O.~Box 500, Batavia, Illinois 60510--5100, USA \\ Email: \texttt{ask@fnal.gov}}}


\begin{abstract}
    Quantum chromodynamics (QCD) reduces the strong interactions, in all their variety, to a simple
    nonabelian gauge theory.
    It clearly and elegantly explains hadrons at short distances, which has led to its universal acceptance.
    Since its advent, however, many of its long-distance, emergent properties have been \emph{believed} to 
    be true, without having been \emph{demonstrated} to be true.
    This article reviews various results in this regime that have been established with lattice gauge 
    theory, directly from the QCD Lagrangian.
    This body of work sheds light on the origin of hadron masses, its interplay with dynamical symmetry 
    breaking, and on other intriguing features such as the phase structure of QCD. 
    Also, nonperturbative QCD is quantitatively important to many aspects of particle physics 
    (especially the quark flavor sector), nuclear physics, and astrophysics.
    This review also surveys some of the most interesting connections to those subjects.
    \vspace*{-12pt}
\end{abstract}

\maketitle

\bibliographystyle{arnuke_revised}

\section{Introduction}

Quantum chromodynamics (QCD) is the modern theory of the strong nuclear force.
It is part of the Standard Model of elementary particles and the underpinning of terrestrial and 
astronomical nuclear physics

The conception of QCD is rightly hailed as a triumph of reductionism, melding the quark model, the idea of
color, and the parton model into a dynamical quantum field theory.
At the same time, the scope of QCD is rich in emergent phenomena.
Symmetries emerge in idealized limits: $C$, $P$, and $T$ are exact when the total ``vacuum angle''
$\bar\theta=0$; chiral symmetries emerge when two or more quark masses
vanish~\cite{Nambu:1960xd,Weinberg:1966fm}; and heavy-quark symmetries are revealed as one or more quark
masses go to infinity~\cite{Shifman:1986sm,Isgur:1989vq}.
More remarkable still are the phenomena that emerge at a dynamically generated energy
scale~$\Lambda_\mathrm{QCD}$, the ``typical scale of QCD.''
Much of what is known about QCD in this nonperturbative regime has long been based on belief.
Evidence from high-energy scattering fostered the opinion that QCD explains the strong interactions and,
therefore, the belief that QCD exhibits certain properties; otherwise, it would not be consistent with
lower-energy observations.
These emergent phenomena---such as chiral symmetry breaking, the generation of hadron masses that are much
larger than the quark masses, and the thermodynamic phase structure---are the most profound phenomena of
gauge theories.
The primary aim of this review is to survey how lattice QCD has enabled us to replace beliefs with knowledge.
To do so, we cover results that are interesting in their own right, influential in a wider arena,
qualitatively noteworthy, and/or quantitatively impressive.

The rest of this article is organized as follows.
Section~\ref{sec:qcd} introduces the QCD Lagrangian and discusses how, in a general setting, to fix its free
parameters.
Section~\ref{sec:lgt} gives a short summary of lattice QCD methodology.
Sections~\ref{sec:spectrum} and~\ref{sec:XSB} discuss hadron masses and their connection to chiral symmetry. 
An output of these calculations are the quark masses and the gauge coupling, which are discussed in
Section~\ref{sec:SM}, along with some timely results pertaining to flavor physics.
Section~\ref{sec:nuke} presents some interesting properties of nucleons.
Section~\ref{sec:thermo} discusses the phase structure of QCD.
Section~\ref{sec:sum} offers some perspective.
The \hyperref[sec:tools]{appendix} identifies resources for readers who wish to start research in numerical
lattice gauge theory.

\section{Quantum Chromodynamics}
\label{sec:qcd}

The (renormalized) Lagrangian of QCD has ``$1+n_f+1$'' free parameters (where $n_f$ is the number of quark 
flavors):
\begin{equation}
    \mathcal{L}_{\mathrm{QCD}} = 
        \frac{1}{2g^2} \tr[F_{\mu\nu}F^{\mu\nu}] -
        \sum_{f=1}^{n_f} \bar{\psi}_f(/ \kern-0.65em D + m_f) \psi_f +
        \frac{i\bar{\theta}}{32\pi^2}\varepsilon^{\mu\nu\rho\sigma}
            \tr[F_{\mu\nu}F_{\rho\sigma}],
    \label{eq:lagrangian}
\end{equation}
where $F^{\mu\nu}$ is the gluon's field strength, $/ \kern-0.65em D=\gamma_\mu(\partial^\mu+A^\mu)$, and
$\psi_f$ denotes the quark field of flavor~$f$.
The first parameter is the gauge coupling~$g^2$; the next $n_f$ parameters are the quark masses~$m_f$; 
and the last parameter, $\bar\theta$, multiplies an interaction that violates \CP\ symmetry.
Experiments have demonstrated the existence of $n_f=6$ quarks.
At energies below the top, bottom, and charm thresholds, however, it is convenient and customary to absorb
the short-distance effects of these quarks into a shift of $g^2$ and then take QCD with $n_f=5$, 4, or~3.
The coupling $g^2$ diminishes gradually with increasing energy, stemming from virtual processes of gluons and
the $n_f$ active quarks; this ``running'' is known as asymptotic freedom~\cite{Politzer:1973fx,Gross:1973id}.

In the Standard Model, quark masses arise from the matrix of Yukawa couplings to the Higgs field, $y$.
The matrix $y$ can be brought into a form with real eigenvalues, $y_f=\sqrt{2}m_f/v$, and an overall phase, 
$\arg\det y$.
(Here, $v=246$~GeV is the Higgs field's vacuum expectation value.)
In this context, the coupling multiplying 
$\varepsilon^{\mu\nu\rho\sigma}\tr[F_{\mu\nu}F_{\rho\sigma}]$ is altered:
$\bar{\theta}=\theta-\arg\det y$, where $\theta$ is allowed in QCD as soon as \CP\ violation is admitted.
Only the difference $\bar{\theta}$ is observable.

Before one can state that a mathematical theory describes or explains the natural world, one must fix its
free parameters with the corresponding number of measurements, in this case $1+n_f+1$.
Because the color of quarks and gluons is confined, the free parameters of QCD must be connected to the
properties of QCD's eigenstates, which are the color-singlet hadrons.
From this perspective, the parameters of QCD may be fixed as follows.
The electric-dipole moment of the neutron is too small to measure, which leads to a bound
$\bar{\theta}<10^{-11}$.
Such delicate cancellation of $\theta$ and $\arg\det y$ is a mystery, known as the strong \CP\
problem~\cite{Kim:2008hd}.
For QCD calculations it means simply that we can set $\bar{\theta}=0$ with no significant consequences.
The rest of the parameters are tuned to reproduce $1+n_f$ specific hadronic properties.
Because the gauge coupling runs, the physical interpretation of $g^2$ is predicated on the energy at which it
reaches a fiducial value, say, $g^2=1$.
But the mathematics is, strictly speaking, dimensionless, so the energy at which $g^2=1$ obtains a physical
meaning as a multiple (or fraction) of some standard mass, such as the proton mass.
In practice, it is wiser to choose the standard mass to be insensitive to the quark masses.
Finally, the $n_f$ quark masses are best related to $n_f$ hadron masses that depend sensitively on them; for
example, the kaon mass is used to tune the strange-quark mass, because $M_K^2\propto m_s$.
With lattice gauge theory~\cite{Wilson:1974sk}, one has a tool to relate the QCD Lagrangian directly to such
hadronic properties and, thereby fix the parameters.
Hadronic properties, by the way, \emph{always} fix the parameters of QCD.
The top quark mass, for example, is measured at the Tevatron via the four-momenta of hadrons in jets.

\section{Numerical Lattice QCD}
\label{sec:lgt}

Lattice gauge theory~\cite{Wilson:1974sk} was invented in an attempt to understand asymptotic freedom 
without gauge-fixing and ghosts~\cite{Wilson:2004de}. 
The key innovation is to formulate nonabelian gauge invariance on a space-time lattice.
Then the functional integrals defining QCD correlation functions are well defined,
\begin{equation}
    \langle\bullet\rangle = \frac{1}{Z}\int \mathcal{D}A\,
        \mathcal{D}\psi\mathcal{D}\bar{\psi}\,
        \, [\bullet] \exp\left(-S\right),
    \label{eq:Z}
\end{equation}
because $\mathcal{D}U$, $\mathcal{D}\psi$, and $\mathcal{D}\bar{\psi}$ are products of a 
countable number of individual differentials. 
Here $S=\int d^4x\,\mathcal{L}_\mathrm{QCD}$ is the action, $\bullet$ is just about any gauge-invariant 
product of fields, and $Z$ ensures that $\langle1\rangle=1$.
This formulation is formally equivalent to classical statistical mechanics, which allows theorists to apply 
a larger tool kit to quantum field theory.
For example, Wilson~\cite{Wilson:1974sk} used a strong-coupling expansion to lowest order in~$1/g^2$ to 
demonstrate confinement. 

The results presented below have been obtained by integrating expressions of the form~(\ref{eq:Z}) on big 
computers with Monte Carlo methods. 
Lattice gauge theory defines QCD mathematically and, thus, in principle provides an algorithm for 
computing anything.
Nevertheless, the computer imposes practical constraints.
To compute anything within a human lifetime, the integrals are defined at imaginary time, $t=-ix_4$, 
which turns Feynman's phase factor into the damped exponential of Equation~\ref{eq:Z}.
A~computer, obviously, has finite memory and processing power, so the spatial volume and time extent of the 
lattice must be finite. 

These limitations do not impair the computation of many important classes of quantities. 
The imaginary time imposes no problem whatsoever for static quantities.
The finite volume introduces errors in one-particle states that are exponentially suppressed and, hence, 
a minor source of uncertainty.
In two-particle states, the finite-volume effects are stronger, \emph{but} the volume dependence 
yields information such as scattering lengths.
Similarly, effects of the finite time extent are exponentially suppressed, except in thermodynamics, 
where it becomes a tool. 
Finally, the continuum limit must be taken as part of the renormalization procedure 
\cite{Symanzik:1983dc,Symanzik:1983gh}.

From Equation~\ref{eq:Z}, it is straightforward to derive some simple results for correlation functions.
The two-point function is
\begin{equation}
    \langle\pi(x_4)\pi^\dagger(0)\rangle = \sum_n 
        |\langle0|\hat{\pi}|\pi_n\rangle|^2 e^{-m_{\pi_n}x_4},
    \label{eq:2pt-m}
\end{equation}
where $\pi$ is a composite field of definite quantum numbers (e.g., of 
the pion), and the sum ranges over all radial excitations.
For a large enough time separation~$x_4$, a fit to an exponential yields the lowest-lying $m_{\pi_1}$ and 
$|\langle0|\hat{\pi}|\pi_1\rangle|$. 
By using a larger set of operators, one can extend this method to compute excited-state properties.
For a transition with no hadrons in the final state, as in leptonic decays, 
one can simply replace $\pi(x_4)$ with a current~$J$,
\begin{equation}
    \langle J(x_4)\pi^\dagger(0)\rangle = \sum_n 
        \langle0|\hat{J}|\pi_n\rangle
        \langle\pi_n|\hat{\pi}^\dagger|0\rangle e^{-m_{\pi_n}x_4}	
    \label{eq:2pt-f}
\end{equation}
in which the only new information is $\langle0|\hat{J}|\pi_n\rangle$, yielding for large $x_4$ the decay 
matrix element of the lowest-lying state. 
For a transition with one hadron in the final state, such as one from a $B$ meson to a pion, one needs a 
three-point function,
\begin{equation}
    \langle\pi(x_4)J(y_4)B^\dagger(0)\rangle = \sum_{mn} 
        \langle0|\hat{\pi}|\pi_m\rangle
        \langle\pi_n|\hat{J}|B_m\rangle
        \langle B_m|\hat{B}^\dagger|0\rangle 
        e^{-m_{\pi_n}(x_4-y_4)-m_{B_m}y_4}
    \label{eq:3pt}
\end{equation}
in which the new information is $\langle\pi_n|\hat{J}|B_m\rangle$, yielding for large $x_4,y_4$ the
matrix element between the lowest-lying states.
We compute matrix elements for flavor-changing processes, dark-matter detection, and nucleon structure in 
this way.

Equations~(\ref{eq:2pt-m})--(\ref{eq:3pt}) are derived by inserting complete sets of eigenstates of the QCD 
Hamiltonian.
The only assumption is that these eigenstates are hadrons.
Thus, every successful fit of these formulae for hadronic correlators provides \emph{a~posteriori}
incre\-mental evidence that hadrons are indeed the eigenstates of QCD.

In all cases of interest, the fermion action is of the form $\bar{\psi}\mathfrak{M}\psi$, where the 
space-time matrix~$\mathfrak{M}$ is a discretization of the Dirac operator (plus quark mass).
Then the fermionic integration in Equation~\ref{eq:Z} can be carried out by hand:
\begin{equation}
    \langle\bullet\rangle = \frac{1}{Z}\int \mathcal{D}A \, [\bullet']
        \det\mathfrak{M} \exp\left(-S_{\mathrm{gauge}}\right).
    \label{eq:Zq}
\end{equation}
Here, the fermionic integration replaces $\psi_i\bar{\psi}_j$ in $\bullet$ with $[\mathfrak{M}^{-1}]_{ij}$ 
to yield~$\bullet'$.
Importance sampling, which is crucial, is feasible only if 
$\det\mathfrak{M}\exp\left(-S_{\mathrm{gauge}}\right)$ is positive.
In most cases, a notable exception being the case of nonzero baryon chemical potential, this condition
holds. 

The determinant $\det\mathfrak{M}$ represents virtual quark-antiquark pairs, also known as sea quarks.
The matrix inverse $\mathfrak{M}^{-1}$ is the propagator of a valence quark moving through a stew of
gluons~$A$ and sea quarks.
Several quark propagators are sewn together to form hadronic correlation functions, which via
Equations~\ref{eq:2pt-m}--\ref{eq:3pt} yield masses and transition matrix elements.
The sea $\det\mathfrak{M}$ poses the largest, and the propagator $\mathfrak{M}^{-1}$ the second largest,
computational challenge.
The numerical algorithms become even more demanding as the quark mass is reduced.
Lattice QCD data with unphysically heavy up and down quarks can be extrapolated to the physical limit guided
by chiral perturbation theory~\cite{Sharpe:2000bc,Bijnens:2007yd}.
This step removes the cloud of unphysically massive pions and replaces it with a better (and improvable)
approximation to the physical pion cloud.

Because $\det\mathfrak{M}$ and $\mathfrak{M}^{-1}$ require so much computing power, several formulations of 
lattice fermions are used. 
As one might anticipate, the computationally fastest and theoretically cleanest methods are not the same.
(If one formulation were both fastest and cleanest, no one would use anything else.)
Each formulation can be characterized by the amount of flavor symmetry retained by the lattice
(Table~\ref{tab:chiral}).
\begin{table}[tbp]
	\centering
	\caption[tab:chiral]{Pattern of chiral symmetry breaking for $n_f$ lattice fermion fields.}
	\begin{tabular}{lr@{$\;\subset\;$}l}
		\hline\hline
		Formulation      & Flavor$\times$space-time & continuum limit \\
		\hline
		Staggered~\cite{Susskind:1976jm,Sharatchandra:1981si} & 
            $\textrm{U}(1)^{n_f}\times\Gamma_4\ltimes\textrm{SW}_4$
		                 & $\textrm{SU}(4n_f)\times\textrm{SU}(4n_f)\times\textrm{SO}(4)$  \\
		Rooted~\cite{Hamber:1983kx} & $\textrm{U}(1)^{n_f}\times\Gamma_4\ltimes\textrm{SW}_4$
		                 & $\textrm{SU}(n_f)\times\textrm{SU}(n_f)\times\textrm{SO}(4)$  \\
		Wilson~\cite{Wilson:1977nj}  & $\textrm{SU}_\mathrm{V}(n_f)\times\textrm{SW}_4$
		                 & $\textrm{SU}(n_f)\times\textrm{SU}(n_f)\times\textrm{SO}(4)$  \\
		Chiral~\cite{Ginsparg:1981bj} & 
            $\textrm{SU}(n_f)\times\textrm{SU}(n_f)\times\textrm{SW}_4$
		                 & $\textrm{SU}(n_f)\times\textrm{SU}(n_f)\times\textrm{SO}(4)$  \\
		\hline\hline
	\end{tabular}
	\label{tab:chiral}
\end{table}
Staggered fermions are computationally the fastest, but the flavor group comes in a semi-direct product with
the symmetry group of the hypercube, SW$_4$, and the total number of species in the continuum limit, for
$n_f$ fields, is $4n_f$.
This fermion doubling is not a problem for the propagators $\mathfrak{M}_\mathrm{stag}^{-1}$.
For the sea, however, one must take~\cite{Hamber:1983kx} $[\det\mathfrak{M}_\mathrm{stag}]^{1/4}$ and appeal
to numerical and perturbative evidence that the rooting yields a local field theory in the continuum 
limit~\cite{Donald:2011if}.

Because of the expense of sea quarks, many lattice QCD calculations have been carried out with two or fewer
(light) flavors.
The error entailed in omitting the strange-quark sea is difficult to estimate, so this review considers
mostly results with 2+1 flavors in the sea.
Here 2+1 denotes the strange sea and two more flavors, for up and down, taken as light as possible.
Such simulations made a breakthrough early in this century~\cite{Davies:2003ik}.
Now the first results with the charmed quark sea, ``2+1+1,'' are becoming available.

\section{Hadron Spectrum}
\label{sec:spectrum}

We compute the masses of hadrons not only for a quantitative comparison of QCD with nature but also to learn
how gauge theories generate mass.
As Section~\ref{sec:SM} shows below, hadron masses are much larger than the sum of the masses of the 
underlying quarks.
The positive binding energy stems from the confining properties of the gluon field and from the kinetic 
energy of the quarks.
\pagebreak

Let us begin with the energy in a flux tube between a static quark--antiquark pair, as a function
of their separation~$\bm{r}$.
The lowest energy level of the flux~tube is the potential energy~$V(r)$, and the excitations of the flux tube
are also informative.
The states are labeled $\Sigma^\pm_{g,u},\Pi_{g,u},\Delta_{g,u},\ldots$, according to the eigenvalues of gluonic
angular momentum along~$\bm{r}$ and of \CP\ [in the subscript $g$ ($u$) stands for 
(\textit{un})\emph{gerade}, which is German for even (odd)].
The $\Sigma$ states also carry a superscript $\pm$ denoting the change of sign (or not) of the wave function
upon reflection in the plane containing $\bm{r}$; otherwise such reflections relate degenerate pairs.

\begin{figure}
    \centering
    \includegraphics[width=0.8\textwidth]{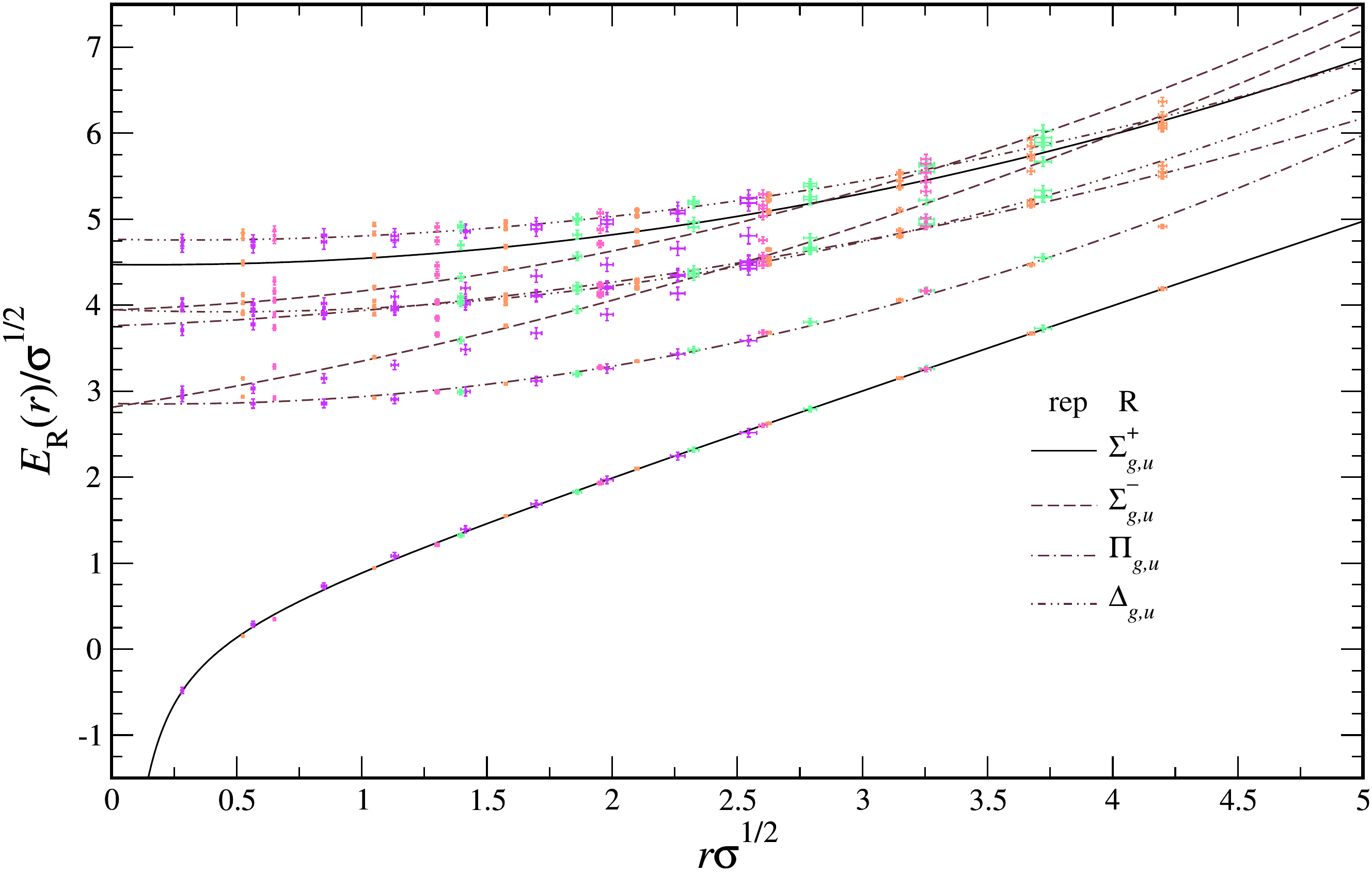}
    \caption[fig:V]{Excitation energies of the chromoelectric field in representation (``rep'')~$R$ between 
        two static sources at separation~$\bm{r}$, in units of the string tension 
        $\sigma\approx(\textrm{400--440~MeV})^2$.
        The lowest energy level $E_{\Sigma_g^+}(r)=V(r)$ is the heavy-quark potential, exhibiting Coulombic 
        behavior at short distances and linearly confining behavior at large distances.
        The higher excitations also exhibit the level ordering of electrodynamics at short distances but 
        the level ordering of a string at large distances.
        The colors pink, orange, green, and violet stand for decreasing lattice spacing.
        Data are from Reference~\citeonline{Juge:2002br}.}
    \label{fig:V}
\end{figure}

Figure~\ref{fig:V} shows the lowest-lying levels in the SU(3) gauge theory without light
quarks~\cite{Juge:2002br}.
At short distances, the level spacing and ordering is consistent with asymptotic freedom: $V(r)$, for
example, is Coulombic up to logarithmic corrections.
As $r$ increases, the spacing changes, and at a separation of around 2~fm, the level ordering rearranges to
that of a string.
The level spacing does not become fully string-like until larger separations~\cite{Juge:2004xr}.
The behavior of the excitations is instructive, because the lowest level, $V(r)$, becomes consistent with a
string at a relatively short distance around $\half$~fm~\cite{Luscher:2002qv}.
A vivid picture of the flux tube has it narrowing as $r$ increases, owing to the attraction between gluons,
but the details suggest that the flux tube looks more like a sausage than a string.

One can imagine connecting the ends of the sausage to obtain non-$q\bar{q}$ states known as glueballs.
Such mesons have no counterpart in the quark model, and lattice gauge theory provides the best (theoretical)
evidence that these states do indeed exist.
Glueball masses with 2+1 flavors of sea quarks show little change~\cite{Richards:2010ck} from earlier
calculations with no sea quarks~\cite{Morningstar:1999rf}.
In particular, the masses remain consistent with the idea, motivated by lattice QCD, that the $f_J(1710)$ is
the lightest scalar glueball~\cite{Sexton:1995kd}.
The pseudoscalar, tensor, and first radially excited scalar glueballs are all 800--900~MeV higher than the
lowest scalar~\cite{Richards:2010ck}.

\begin{figure}
    \centering
    \includegraphics[width=0.8\textwidth]{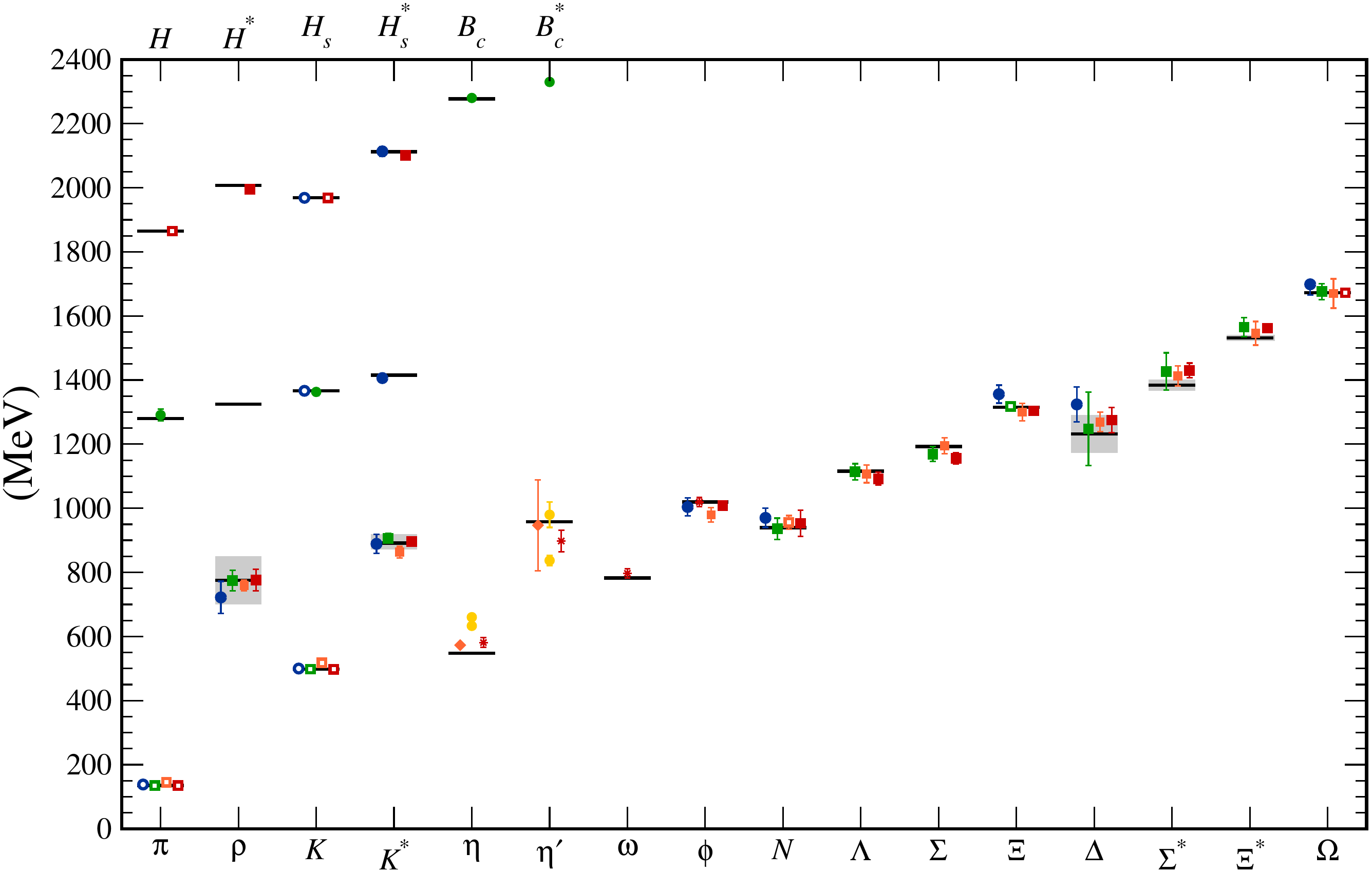}
    \caption[fig:spectrum]{Hadron spectrum from lattice QCD.
        Comprehensive results for mesons and baryons are from 
        MILC~\cite{Aubin:2004wf,Bazavov:2009bb},
        PACS-CS~\cite{Aoki:2008sm},
        BMW~\cite{Durr:2008zz}, and
        QCDSF~\cite{Bietenholz:2011qq}.
        Results for $\eta$ and $\eta'$ are from 
        RBC \& UKQCD~\cite{Christ:2010dd},
        Hadron Spectrum~\cite{Dudek:2011tt} (also the only $\omega$ mass), and
        UKQCD~\cite{Gregory:2011sg}.
        Results for heavy-light hadrons from 
        Fermilab-MILC~\cite{Bernard:2010fr},
        HPQCD~\cite{Gregory:2010gm}, and
        Mohler \& Woloshyn~\cite{Mohler:2011ke}.
        Circles, squares, and diamonds stand for staggered, Wilson, and chiral sea quarks, respectively. 
        Asterisks represent anisotropic lattices.
        Open symbols denote the masses used to fix parameters. 
        Filled symbols (and asterisks) denote results.
        Red, orange, yellow, green, and blue stand for increasing numbers of ensembles 
        (i.e., lattice spacing and sea quark mass).
        Horizontal bars (gray boxes) denote experimentally measured masses (widths).
        $b$-flavored meson masses are offset by $-4000$~MeV.}
    \label{fig:spectrum}
\end{figure}

Lattice QCD has been used to verify the mass spectrum of quark-model hadrons within a few percent.
Figure~\ref{fig:spectrum} shows four broad efforts on the spectrum of the isopsin-1 light mesons and the
isospin-\half\ and -\case{3}{2}
baryons~\cite{Aubin:2004wf,Bazavov:2009bb,Aoki:2008sm,Durr:2008zz,Bietenholz:2011qq}.
All these simulations include $2+1$ flavors of sea quarks, and the error bars in
References~\citeonline{Aubin:2004wf,Bazavov:2009bb,Durr:2008zz} reflect thorough analyses of the systematic 
uncertainties.
A~satisfying feature of Figure~\ref{fig:spectrum} is that the results do not depend in a systematic way on 
the fermion formulation chosen for the quarks.
Even the latest results for the difficult $\eta$-$\eta'$ splitting are 
encouraging~\cite{Christ:2010dd,Dudek:2011tt,Gregory:2011sg}.

Figure~\ref{fig:spectrum} includes predictions for mesons with quark content~$\bar{b}c$
\cite{Allison:2004be,Gregory:2010gm,Gregory:2009hq}.
The prediction for the pseudoscalar $B_c$ has been (subsequently) confirmed by 
experiment~\cite{Abulencia:2005usa,Aaltonen:2007gv},
whereas the prediction for the vector $B_c^*$ awaits confirmation.
These predictions build on successful calculations of the $b\bar{b}$ and $c\bar{c}$ spectra 
\cite{Mohler:2011ke,Gray:2005ur,Meinel:2009rd,Burch:2009az,Dowdall:2011wh}, which reproduce 
the experimental results well.

The most striking aspect of the spectrum is how well it agrees with nature.
The nucleons provide almost all the mass in everyday objects, and their masses have been verified 
within 3.5\%.
Their mass mostly comes, via $m=E/c^2$, from the kinetic energy of the quarks and the energy stored in the
sausage-like flux tube(s) holding the quarks together.

\section{Chiral Symmetry Breaking}
\label{sec:XSB} 

A~striking feature of the hadron spectrum in Figure~\ref{fig:spectrum} is that the pion has a small mass,
around 135~MeV, whereas the other hadrons have masses more than five times larger.
To understand the origin of the difference, Nambu~\cite{Nambu:1960xd} applied lessons from superconductivity,
noting (four years before quarks were proposed) that a spontaneously broken axial symmetry would constrain
the pion's mass to vanish, with small explicit symmetry breaking allowing it to be nonzero.

If the up and down quarks can be neglected, the QCD Lagrangian acquires an
$\mathrm{SU}_{\mathrm{L}}(2)\times\mathrm{SU}_{\mathrm{R}}(2)$ symmetry, thereby explaining the origin of
Nambu's axial symmetry.
The consequences of spontaneous symmetry breaking were studied further by Goldstone~\cite{Goldstone:1961eq},
which led to the formula~\cite{Goldstone:1962es},
\begin{equation}
    m_\pi^2 \langle\bar{\psi}\psi\rangle = 0,
    \label{eq:goldstone}
\end{equation}
when applied to QCD with massless up and down quarks.
The flavor-singlet vacuum expectation value $\langle\bar{\psi}\psi\rangle$ is known as the chiral condensate.
If either factor on the left-hand side of Equation~\ref{eq:goldstone} is nonzero, then the other must vanish.

From the earliest days of QCD, most physicists were confident that this theory explained the richness of the 
strong interactions.
Because both QCD and Nambu's picture of the pion were considered correct, it was believed that QCD \emph{must}
generate a chiral condensate.
Until recently, however, no \emph{ab initio} calculation of $\langle\bar{\psi}\psi\rangle$ tested
Equation~\ref{eq:goldstone}.
Lattice QCD has now filled this gap~\cite{DeGrand:2006nv,Fukaya:2009fh,Fukaya:2010na}:
\begin{equation}
    \langle\bar{\psi}\psi\rangle = 
        \left[234\pm4\pm17~\mathrm{MeV}\right]^3 \quad
        \mbox{($\overline\mathrm{MS}$ scheme at 2~GeV)}.
    \label{eq:condensate}
\end{equation}
Here, the first uncertainty is statistical and the second is a combination of systematics, and the quark 
masses have been adjusted to Nambu's idealization, $\mup=\mdn\to0$, $\mstr$~physical~\cite{Fukaya:2010na}.
In summary, QCD's symmetries and dynamics have now been demonstrated to account for the hierarchy 
between the pion and the other hadron masses.

\section{Standard Model Parameters}
\label{sec:SM}

The Standard Model (with nonzero neutrino masses and mixing angles) has 28 free parameters:
\begin{itemize}
    \item Gauge couplings: $\alphas$, $\alpha_\mathrm{QED}$, $\alpha_\mathrm{W}=(M_\mathrm{W}/v)^2/\pi$;
     \item Quark sector: $\mup e^{i\bar{\theta}}$, $\mdn$, $\mstr$, $\mch$, $\mbt$, \mtop;
         $|V_\mathrm{us}|$, $|V_\mathrm{cb}|$, $|V_\mathrm{ub}|$, $\delta_\mathrm{KM}$;
    \item Lepton sector: $m_{\nu_1}$, $m_{\nu_2}$, $m_{\nu_3}$,
         $m_\mathrm{e}$, $m_\mu$, $m_\tau$;
         $\theta_{12}$, $\theta_{23}$, $\theta_{13}$, $\delta_\mathrm{PMNS}$,
         $\alpha_{21}$, $\alpha_{31}$;
     \item Standard electroweak symmetry breaking: $v=246$~GeV, 
         $\lambda=(M_{H}/v)^2/2$.
\end{itemize}
Lattice QCD is essential or important in determining the values of eleven parameters (the first under gauge 
couplings and all but $\mtop$ under quark sector). 
Lattice field theory (without QCD) is also useful for shedding light on the Higgs
self-coupling~$\lambda$~\cite{Gerhold:2010bh} and the top Yukawa coupling
$y_{\mathrm{t}}=\sqrt{2}\mtop/v$~\cite{Gerhold:2009ub}.

As in Equation~\ref{eq:condensate}, this section reports several results in the conventional
$\overline\mathrm{MS}$ scheme at a renormalization point $\mu$, such as 2~GeV or the $Z$ boson mass $M_Z$.
$\overline\mathrm{MS}$~is the most common renormalization scheme in quantum field theory, because it is the 
technically simplest one in perturbative calculations.
Readers unfamiliar with the $\overline\mathrm{MS}$ scheme should focus on the consistency and on the 
relative uncertainty of the results, rather than the details of the scheme's definition.

\subsection{Quark masses and \boldmath\alphas}

Confinement precludes the direct measurement of quark masses.
Instead, the masses in Equation~\ref{eq:lagrangian} must be determined from closely associated measurable 
properties of hadrons.
The $n_f$ bare quark masses are adjusted until $n_f$ hadron masses of suitable flavor agree with 
experimental measurements.
Table~\ref{tab:q} shows four sets of results, following conversion to the $\overline\mathrm{MS}$ scheme 
described above.
\begin{table}
    \centering
    \caption[tab:q]{Quark masses from lattice QCD converted to the $\overline\mathrm{MS}$ scheme and run to 
        the scale indicated.
        Entries are in MeV.}
    \label{tab:q}
    \vspace*{3pt}
    \begin{tabular}{r|*{5}{r@{$\,\pm\,$}l}}
        \hline\hline
         Flavor (scale) & 
         \multicolumn{2}{c}{Ref.~\cite{Bazavov:2009bb}} &
         \multicolumn{2}{c}{Ref.~\cite{Davies:2009ih}} &
         \multicolumn{2}{c}{Ref.~\cite{Blum:2010ym}} &
         \multicolumn{2}{c}{Ref.~\cite{Durr:2010vn}} &
         \multicolumn{2}{c}{Ref.~\cite{McNeile:2010ji}} \\
        \hline
        $\Mup(2~\textrm{GeV})$ & 1.9&0.2 & 2.01&0.14 & 2.24&0.35 & 2.15&0.11 \\
        $\Mdn(2~\textrm{GeV})$ & 4.6&0.3 & 4.79&0.16 & 4.65&0.35 & 4.79&0.14 \\
        $\Mstr(2~\textrm{GeV})$ &  88&5  & 92.4&1.5  & 97.7&6.2  & 95.5&1.9 \\
        $\Mch(3~\textrm{GeV})$ & \multicolumn{2}{c}{} & \multicolumn{2}{c}{} & 
            \multicolumn{2}{c}{}& \multicolumn{2}{c}{} & 986&10 \\
        $\Mbt(10~\textrm{GeV})$ & \multicolumn{2}{c}{} & 
            \multicolumn{2}{c}{}& \multicolumn{2}{c}{} & \multicolumn{2}{c}{}  & 3617&25 \\
        \hline\hline
    \end{tabular}
\end{table}
The results in the first, third, and fourth columns are completely independent, employing different
formulations for sea quarks and different treatments of electromagnetic effects.
The results in the second column are derived from mass ratios [$2\mstr/(\mdn+\mup)=27.3\pm0.3$ and
$\mup/\mdn=0.42\pm0.04$] underlying those in the first column, combined with precise values of the ratio
$\mch/\mstr=11.85\pm0.16$~\cite{Davies:2009ih} and $\bar{m}_c$~\cite{Allison:2008xk}.

These results are remarkable for at least two reasons.
First, the up and down masses are very small, approximately four and nine times the electron mass, 
respectively.
Quark masses arise from interactions with the Higgs field (or something like it).
Thus, this sector is not the origin of much of the mass of everyday objects.
Second, $\mup$, although very small, is nevertheless significantly far from zero.
This outcome is interesting because if $\mup=0$, then the additional symmetry of the Lagrangian would 
render $\bar{\theta}$ unphysical, obviating the strong \CP\ problem~\cite{Kim:2008hd}.
(A subtlety could arise from a nonperturbative additive correction to \mup, but it is probably too small to 
alter this conclusion.)

The heavy charmed quark and bottom quark masses can be determined along the same lines, but the most 
accurate results come from computing quarkonium correlation functions and taking their continuum 
limit~\cite{Bochkarev:1995ai}.
These functions give spacelike information on the same function measured in the timelike region in 
$e^+e^-\to c\bar{c}$ or $e^+e^-\to b\bar{b}$.
Perturbation theory to order $\alphas^3$~\cite{Chetyrkin:2006xg} can then be used to extract the 
heavy-quark masses and $\alphas$~\cite{Allison:2008xk}.
This approach yields the results in the fifth column of Table~\ref{tab:q}, which are in nearly perfect 
agreement with the corresponding determinations from $e^+e^-$ collisions~\cite{Chetyrkin:2009fv}.

Lattice QCD provides excellent ways to determine the gauge coupling~$\alphas=g^2/4\pi$.
In lattice gauge theory, the bare coupling $g_0^2$ is an input.
Alas, for many lattice gauge actions, perturbation theory in $g_0^2$ converges poorly~\cite{Lepage:1992xa},
obstructing a perturbative conversion to the $\overline\mathrm{MS}$ or other such schemes.
Two other strategies have been adopted to circumvent this obstacle.
One is to compute a short-distance lattice quantity---such as a Wilson loop---and reexpress perturbation 
theory for the Monte Carlo results in a way that eliminates $g_0^2$ in favor of a renormalized
$g^2$~\cite{Lepage:1992xa,Mason:2005zx}.
The other is to compute a short-distance quantity with a continuum limit, and then apply continuum
perturbation theory.
The quarkonium correlator used for $\mch$ and $\mbt$ is an example~\cite{Chetyrkin:2006xg}.
Other examples include the Schr\"odinger functional~\cite{Luscher:1992an} and the Adler
function~\cite{Eidelman:1998vc,Baikov:2008jh}.

Results from several complementary lattice QCD methods \cite{Allison:2008xk,Davies:2008sw,Aoki:2009tf,%
Shintani:2010ph,Blossier:2012ef} are collected in Table~\ref{tab:alpha} and compared with an average of 
determinations from high-energy scattering and decays~\cite{Bethke:2009jm}.
\begin{table}
    \centering
    \caption[tab:alpha]{Values of $\alphas(M_Z)$ (in the $\overline\mathrm{MS}$ scheme) from lattice QCD and 
        an average of determinations from high-energy scattering and decays.
        An update to the values in the first two rows can be found in Reference~\citeonline{McNeile:2010ji}.
        The central values and error bars from References~\citeonline{Aoki:2009tf,Shintani:2010ph} have been 
        symmetrized to ease comparison.
        In principle, ``nature's sea'' includes non--Standard Model colored particles.}
    \label{tab:alpha}
    \vspace*{3pt}
    \begin{tabular}{r@{$\,\pm\,$}llll}
        \hline\hline
        \multicolumn{2}{c}{$\alphas(M_Z)$} & Observable & 
            Sea formulation  & Ref. \\
        \hline
        0.1174&0.0012 & Charmonium correlator & 
            2+1 asqtad staggered & \cite{Allison:2008xk} \\
        0.1183&0.0008 & Small Wilson loops &
            2+1 asqtad staggered & \cite{Davies:2008sw} \\
        0.1197&0.0013 & Schr\"odinger functional & 
            2+1 improved Wilson & \cite{Aoki:2009tf} \\
        0.1185&0.0009 & Adler function & 
            2+1 overlap & \cite{Shintani:2010ph} \\
        0.1200&0.0014 & Ghost-gluon vertex & 
            2+1+1 twisted Wilson & \cite{Blossier:2012ef} \\
        \hline
        0.1186&0.0011 & Scattering, $\tau$ decay, etc.\ & 
            Nature's sea &  \cite{Bethke:2009jm} \\
        \hline\hline
    \end{tabular}
\end{table}
There is excellent consistency among the results, not only with different discretizations of the determinant
for sea quarks but also when the charmed sea is included~\cite{Blossier:2012ef}.
An important source of uncertainty is the truncation of perturbation theory, including strategies for
matching to the $\overline\mathrm{MS}$ scheme, and running to scale $M_Z$.
In the example of the small Wilson loops, an independent analysis of the data from
Reference~\citeonline{Davies:2008sw} has been carried out; this analysis found
$\alphas(M_Z)=0.1192\pm0.011$~\cite{Maltman:2008bx}, which should be compared with the second line of
Table~\ref{tab:alpha}.

The agreement between the lattice QCD and perturbative QCD results for $\alphas$, $\mch$, and $\mbt$ is 
especially compelling because QCD is the union of the quark model of hadrons and the parton model of 
high-energy scattering.
This consistency is evidence that the QCD of hadrons and the QCD of partons are the same.

\subsection{Flavor Physics}
\label{sec:ckm}

As mentioned in Section~\ref{sec:qcd}, the quark masses arise from the electroweak interactions.
In a basis where the mass matrix is diagonal, the $W$ boso couples to all combinations of
$(u,c,t,\ldots)\otimes(d,s,b,\ldots)^\mathrm{T}$ quarks.
Along with the SU(2) gauge coupling, the $udW$ vertex carries a factor $V_{ud}$, and similarly for all other
combinations.
As a change of basis (from the weak interaction basis to the mass basis), the
Cabibbo--Kobayashi--Maskawa (CKM)~\cite{Cabibbo:1963yz,Kobayashi:1973fv} quark-mixing matrix~$V$ is unitary.
After considering global symmetries of the gauge interactions, one sees that the CKM matrix has fewer
parameters than a generic unitary matrix does.
For three generations, three mixing angles and one \CP-violating phase remain, and a convenient choice 
consists of $|V_{us}|$, $|V_{ub}|$, $|V_{cb}|$, and $\arg V_{ub}^*$.

Lattice QCD calculations play a key role in flavor physics.
The phase and, except for $|V_{tb}|$, all magnitudes of the CKM matrix can be accessed via processes for 
which the corresponding lattice-QCD calculations are under good control.
(Other processes that do not need lattice QCD also provide information on the CKM matrix.)
A representative set of calculations and their utility is illustrated~by
\begin{equation}
V = \left( \begin{array}{cccc}
      |V_{ud}|  & |V_{us}|    & |V_{ub}| & \,\arg V_{ub}^* \\
    \mbox{\small$\pi\to\ell\nu$}  & \mbox{\small$K\to\ell\nu$}    & \mbox{\small$B\to\tau\nu$} &
        \mbox{\small$\langle K^0|\bar{K}^0\rangle$} \\
  \mbox{\small$n\to pe^-\bar\nu$} & \mbox{\small$K\to\pi\ell\nu$} & \mbox{\small$B\to\pi\ell\nu$} & \\[0.5em]
      |V_{cd}|  & |V_{cs}|    & |V_{cb}|   \\
     \mbox{\small$D\to\ell\nu$}   & \mbox{\small$D_s\to\ell\nu$}  & \mbox{\small$B\to D\ell\nu$} \\
    \mbox{\small$D\to\pi\ell\nu$} & \mbox{\small$D\to K\ell\nu$}  & \mbox{\small$B\to D^*\ell\nu$}  \\[0.5em]
      |V_{td}|  & |V_{ts}|    & |V_{tb}|   \\
    \mbox{\small$\langle B_d|\bar{B_d}\rangle$}  & \mbox{\small$\langle B_s|\bar{B}_s\rangle$} & 
    \mbox{\small(no $t\bar q$ hadrons)}
\end{array} \right).
\end{equation}
These leptonic and semileptonic decays (first two rows) or meson-antimeson oscillations (phase and third row)
have one hadron in the initial state and one (or none) in the final state.
Thus, all of these flavor-changing amplitudes can be computed in lattice QCD via Equations~\ref{eq:2pt-f}
or~\ref{eq:3pt}.
Semileptonic transition form factors for $K\to\pi\ell\nu$~\cite{Lubicz:2009ht,Boyle:2010bh},
$B\to\pi\ell\nu$~\cite{Dalgic:2006dt,Bailey:2008wp}, and $B\to D^*\ell\nu$~\cite{Bernard:2008dn} are
sensitive to the mixing angles, and $K^0$-$\bar K^0$
mixing~\cite{Aubin:2009jh,Aoki:2010pe,Durr:2011ap,Bae:2011ff} is sensitive to the \CP-violating phase.
Together with calculations of $D$ meson decays~\cite{Aubin:2004ej,Na:2010uf,Na:2011mc} and 
$B^0_{(s)}$-$\bar B^0_{(s)}$ mixing~\cite{Gamiz:2009ku}, the full suite of lattice QCD calculations
overdetermines the CKM matrix and, thus, tests for consistency.
The semileptonic $D$ decays are considered crosschecks.
Taking $|V_{cd}|$ and $|V_{cs}|$ from CKM unitarity (which is very precise), one finds that 
lattice QCD calculations of the kinematic distributions~\cite{Aubin:2004ej} and the
normalization of the rate~\cite{Na:2010uf,Na:2011mc} agree with results from several experiments.

Non-SM particles could spoil this picture, which is why it is interesting to test it in detail.
With a fourth generation of quarks and leptons, the $3\times3$ submatrix generically would not be unitary.
If other particles, such as supersymmetric partners of the known particles, change quark flavor, then the SM
relation between a flavor-changing process and $V$ is spoiled.
The off-diagonal elements are small---$|V_{us}|\sim0.2$, $|V_{cb}|\sim4\times10\,^{-2}$, and
$|V_{ub}|\sim3\times10\,^{-3}$---so it is not out of the question for (widely anticipated) TeV-scale
particles to make detectable contributions.

During the first decade of the twenty-first century, all simple lattice-QCD calculations that are pertinent
to the CKM matrix were carried out with 2+1 sea quarks.
In most cases, more than one collaboration has published results, and, in many cases, the literature covers
more than one fermion formulation for the quarks.
The calculations most directly connected to determining the CKM parameters have been combined---with an eye
to correlations in the errors---in Reference~\citeonline{Laiho:2009eu}.
(Updates are available at \url{http://latticeaverages.org/}).

Despite the broad agreement between flavor-physics measurements and the Standard Model, some tension appears 
in global fits to the four CKM parameters~\cite{Lunghi:2008aa}.
Some mild discrepancies also arise in a few isolated processes, and here we consider two leptonic
decays in which lattice QCD plays a key role.

Let us begin by noting that the semileptonic and leptonic determinations of $|V_{us}|$, which rely on
the matrix elements of References~\citeonline{Lubicz:2009ht,Boyle:2010bh} and~%
\citeonline{Follana:2007uv,Durr:2010hr,Bazavov:2010hj}, respectively, are completely
compatible~\cite{Colangelo:2010et}.
The vector component of the $W$ boson mediates the former, and the axial current mediates the latter.
Because these determinations are compatible, nothing other than the SM $W$ boson with its $V-A$ coupling is
needed to account for these decays.

The present status of semileptonic and leptonic determinations of $|V_{ub}|$ is not so tidy.
Combining lattice QCD for the $B\to\pi\ell\nu$ form factor~\cite{Dalgic:2006dt,Bailey:2008wp} with
measurements from BaBar~\cite{:2010uj} yields $|V_{ub}|=(2.95\pm0.31)\times10^{-3}$; with
Belle~\cite{Ha:2010rf}, $|V_{ub}|=(3.43\pm0.33)\times10^{-3}$.
The average taken here is $|V_{ub}|_{B\to\pi\ell\nu}=(3.19\pm0.32)\times10^{-3}$.
Combining lattice QCD for the $B$ meson decay constant \cite{Gamiz:2009ku,Bazavov:2011aa}
with the world average of the rate for $B^+\to\tau^+\nu$~\cite{Asner:2010qj}, 
however, suggests that $|V_{ub}|_{B\to\tau\nu}=(4.95\pm0.55)\times10^{-3}$, which deviates by $2.8\sigma$ 
from $|V_{ub}|_{B\to\pi\ell\nu}$.
This discrepancy could be explained if another particle (or particles) were to mediate the decays with a 
coupling different from the $W$ boson's~$V-A$.
Examples include a charged Higgs boson~\cite{Akeroyd:2007eh,Deschamps:2009rh} and a right-handed vector 
current~\cite{Crivellin:2009sd}.
The plot thickens when one considers inclusive charmless semileptonic $B$ decays, which are mediated by all 
possible currents.
These decays imply a value of $|V_{ub}|$ in between those from the two exclusive methods.

\begin{figure}
    \vspace*{-8pt}
    \centering
    \includegraphics[width=0.8\textwidth]{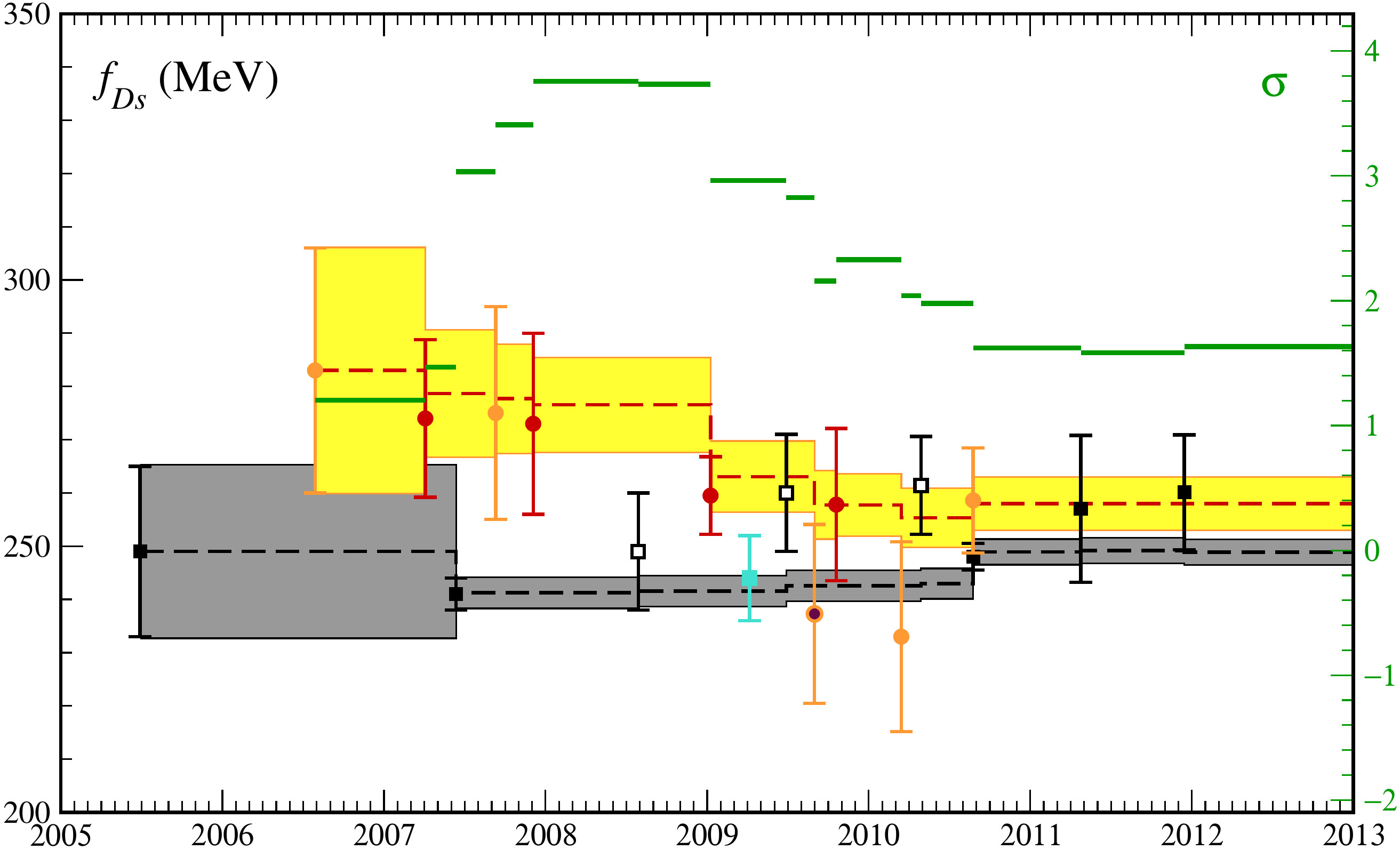}
    \vspace*{-4pt}
    \caption[fig:fDs]{Comparison between values for $f_{D_s}$ since 2005.
        The green line (\emph{right axis}) shows the discrepancy in~$\sigma$.
        The black line shows the running average (with $\pm1\sigma$ error band in gray) of 2+1-flavor 
        lattice QCD calculations from Fermilab-MILC~\cite{Aubin:2005ar,Bazavov:2011aa}, 
        HPQCD~\cite{Follana:2007uv,Davies:2010ip}, and PACS-CS \cite{Namekawa:2011wt}.
        Black filled (open) points stand for published (preliminary) results.
        [The two-flavor cyan point from ETM~\cite{Blossier:2009bx} is not included in the average.]
        The red line shows the running average (with $\pm1\sigma$ error band in yellow) of measurements from 
        BaBar, Belle, and CLEO-$c$~\cite{Asner:2010qj}.
        Orange (red) points stand for results from BaBar and Belle (CLEO-$c$).
        Updated from Reference~\citeonline{Kronfeld:2009cf}.}
    \label{fig:fDs}
\end{figure}

The history of results on the $D_s\to\mu^+\nu$ and $D_s\to\tau^+\nu$ decays suggests caution, however.
In 2008, the measured branching fractions exceeded the SM predictions by nearly $4\sigma$ 
(Figure~\ref{fig:fDs}).
This discrepancy relies on the decay constant $f_{D_s}$ from lattice QCD~\cite{Aubin:2005ar,Follana:2007uv}.
Through the use of the same methods as for the $\pi$ and $K$ decay 
constants~\cite{Follana:2007uv,Davies:2010ip}, the $D_s$ decay constant can be computed to 1--2\%.
At the time, some experimenters asserted that either the lattice QCD calculations were wrong or that New 
Physics mediated the decay.
For example, the excesses of $D_s\to\ell^+\nu$ could be due to leptoquarks~\cite{Dobrescu:2008er};
a few-percent amplitude could be constructively interfering.
(Leptoquarks are hypothetical particles with lepton and quark quantum numbers.)
As more measurements came in, however, the discrepancy softened, and it is now only $1.6\sigma$.
Although the lattice QCD average of the $D_s$ decay constant has increased by 8~MeV,
the experimental average has decreased by 18~MeV.
The early measurements fluctuated upward; perhaps the same holds for $B\to\tau\nu$.

\section{Nucleon Matrix Elements, Dark Matter, and the LHC}
\label{sec:nuke}

Two of the most compelling questions facing particle physics are the origin of electroweak symmetry breaking 
and the composition of dark matter.
The experiments developed to address them rely on the familiar proton and neutron.
The Large Hadron Collider (LHC) collides $pp$, and the Tevatron $p\bar p$, and detectors buried deep 
underground hope to observe weakly interacting massive particles (WIMPs) scatter of the protons and
neutrons in nuclei.
To interpret the results of these experiments, it is helpful to calculate certain matrix elements of the 
nucleon~\cite{Bhattacharya:2011qm}.

Let us start with WIMP-nucleon scattering.
Each quark's contribution to the cross section is proportional to a matrix element known as the sigma term,
\begin{equation}
    \sigma_q = m_q\langle N|\bar{q}q|N\rangle = m_q\frac{\partial M_N}{\partial m_q},
    \label{eq:Sig-q}
\end{equation}
where $m_q$ is the quark mass, and $M_N$ is the nucleon mass.
All quarks $q$ in the nucleon can contribute, including virtual quarks such as $s$ and~$c$.
The partial derivative should be taken with the other $n_f$ QCD parameters held fixed.
Usually the light quarks are combined into the isospin-singlet 
\begin{equation}
    \sigma_{u+d}=\half(\mup+\mdn)\langle N|(\bar{u}u+\bar{d}d)|N\rangle.
    \label{eq:Sig-ud}
\end{equation}
(Beware of factors of 2 in the definitions of these and similar quantities, sometimes 
denoted $\Sigma$, in the literature.)
For WIMP detection, $\sigma_{u+d}$ and $\sigma_s$ are especially important, because virtual pairs of heavier 
quarks are far too uncommon inside nucleons.

Until recent lattice QCD calculations became available, the light quark sigma term was extracted from $\pi N$
scattering, with the help of chiral perturbation theory ($\chi$PT)~\cite{Koch:1982pu,Pavan:2001wz}.
As shown in the first two rows of the $\sigma_{u+d}$ column of Table~\ref{tab:sigma}, the extraction depends
more on assumptions than on the experimental data.
To estimate $\sigma_s$, one uses information from the baryon octet masses~\cite{Borasoy:1996bx}.
Unfortunately, this information must be subtracted from $\sigma_{u+d}$, which renders $\sigma_s$ rather
unstable.
There is a pressing need to improve both matrix elements~\cite{Ellis:2009ai}.

\begin{table}
    \centering
    \caption[tab:sigma]{Table of scalar-density matrix elements.
        The first and fourth rows use 2 flavors of sea quarks; the others use 2+1.
        Entries are in MeV.}
    \label{tab:sigma}
    \vspace{3pt}
    \begin{tabular}{lccr}
        \hline\hline
        Method & $\sigma_{u+d}$ & $\sigma_s$ & Reference \\
        \hline
        $\pi N$ scattering $\oplus$ baryon octet masses & $45 \pm 8$ & $122 \pm 143$ & 
            \cite{Koch:1982pu,Borasoy:1996bx} \\
        & $64 \pm 7$ & $378 \pm 135$ & \cite{Pavan:2001wz,Borasoy:1996bx} \\
        \hline
        Lattice QCD Feynman--Hellmann & $53\pm2^{+21}_{-\,7}$ &  & \cite{Ohki:2008ff} \\ 
        Lattice QCD $M_N$ $\chi$PT  & $47\pm8\pm3$ & $31\pm15\pm4$ & \cite{Young:2009zb} \\ 
        Lattice QCD Feynman--Hellmann &  & $59\pm7\pm8$ & \cite{Toussaint:2009pz} \\ 
        Lattice QCD matrix element  &  & $30\pm8\pm21$ & \cite{Takeda:2010cw} \\ 
        Lattice QCD Feynman--Hellmann & $39\pm4^{+18}_{-\,7}$ & $34\pm14^{+28}_{-23}$ & \cite{Durr:2011mp} \\
        Lattice QCD Feynman--Hellmann & $31\pm3\pm4$ & $71\pm34\pm59$ & 
            \cite{Horsley:2011wr} \\ 
        \hline\hline
    \end{tabular}
\end{table}

Equation~\ref{eq:Sig-q} suggests two methods to compute $\sigma_q$ in lattice QCD: either from a three-point
function, as in Equation~\ref{eq:3pt}, or through study of the mass dependence of the nucleon mass~$M_N$.
The latter is known as the Feynman--Hellmann theorem, and here one can either reweight the Monte Carlo
ensemble to take the derivative locally or study the chiral extrapolation to obtain a global handle on the
derivative.
Table~\ref{tab:sigma} compiles several recent results.
From a quantitative perspective, it seems that the results are still settling down, although they tend to
favor lower values of $\sigma_{u+d}$.
More strikingly (and self-consistently with low $\sigma_{u+d}$), the results for~$\sigma_s$ substantially
contradict the high values used in WIMP phenomenology.
Even if these results are not yet as mature as those reported in Sections~\ref{sec:spectrum}, \ref{sec:XSB},
and~\ref{sec:SM}, they seem to give a more stable picture than the non-lattice estimates.
Further work and promising new methods~\cite{Dinter:2012tt} should clarify the sigma terms soon.

\begin{figure}[t]
    \centering
    \includegraphics[width=0.8\textwidth]{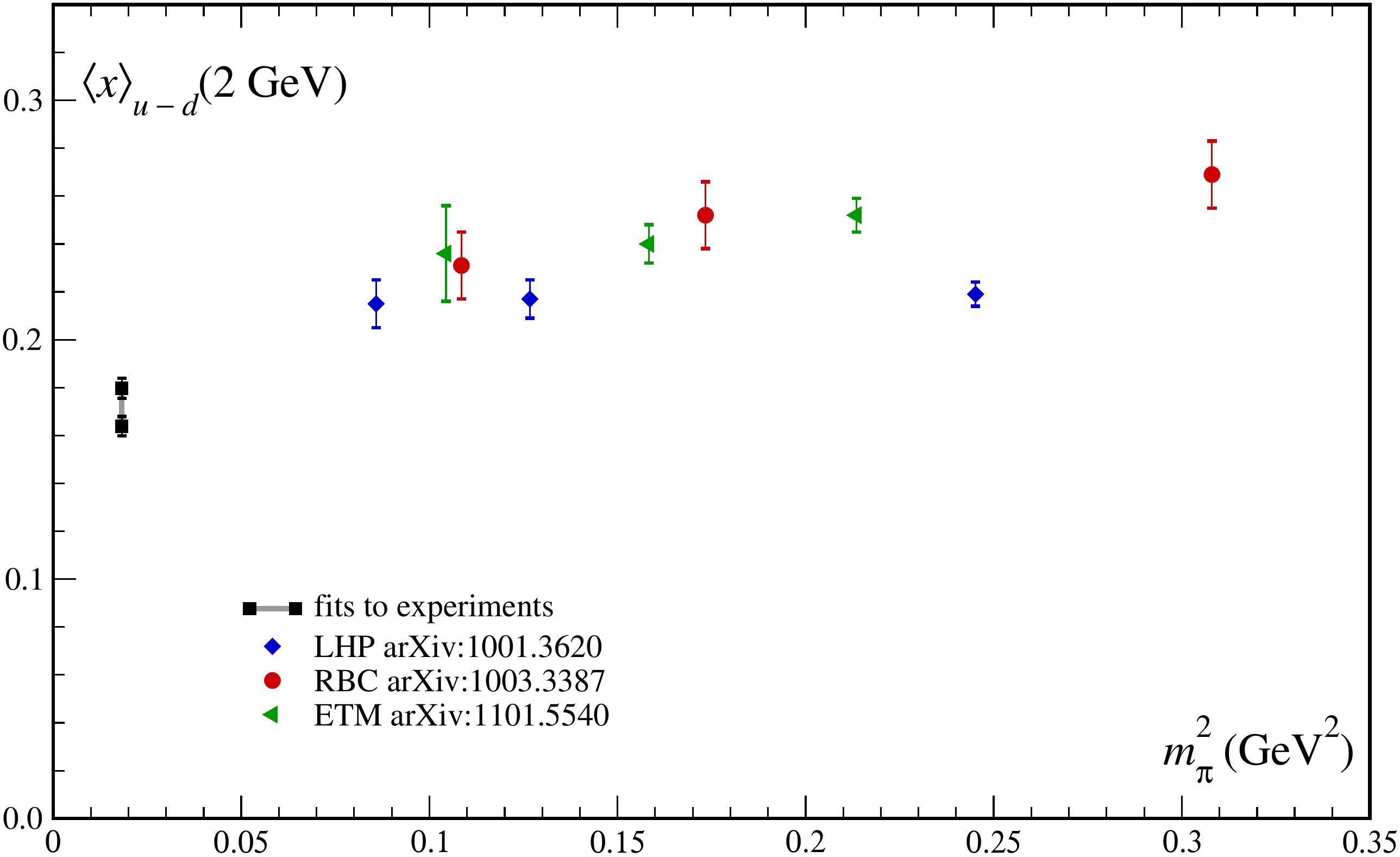}
    \caption[fig:x]{Nonsinglet average momentum fraction $\langle x\rangle_{u-d}$ vs.\ $m_\pi^2$ from
        LHP~\cite{Bratt:2010jn}, RBC \& UKQCD~\cite{Aoki:2010xg}, and ETM~\cite{Dinter:2011jt}.
        The last has 2+1+1 flavors of sea quarks, the others have 2+1 flavors.
        The black squares show two fits~\cite{Martin:2009iq,Alekhin:2012ig} to experimental data; 
        other recent fits of this kind fall between these two~\cite{Alekhin:2012ig}.} 
    \label{fig:x}
\end{figure}

In $pp$ or $p\bar p$ collisions, the essential long-distance ingredient in computing cross sections are the
moments of the parton distribution functions.
These moments are given by matrix elements of local operators, similar to the $\bar qq$ in
Equation~\ref{eq:Sig-q} but with different Dirac structures, such as $\gamma^\mu$ and $\gamma^\mu\gamma^5$,
and derivatives to pull out higher powers of the momentum fraction.
Figure~\ref{fig:x} shows some recent lattice QCD results for the nonsinglet average momentum fraction
$\langle x\rangle_{u-d}$ as a function of simulation $m_\pi^2$~\cite{Bratt:2010jn,Aoki:2010xg,Dinter:2011jt},
compared with two phenomenological results~\cite{Martin:2009iq,Alekhin:2012ig}.
The latter are obtained by fitting experimental data, which exist over a large but limited range of $x$, to
reasonable parameterizations.
In principle, the lattice QCD moments add extra information, but the status of the chiral extrapolation may
preclude this step at this time, even though some functional forms lead to
results~\cite{Bratt:2010jn,Aoki:2010xg} that agree with the fits to experiment.
In this regard, earlier work with 2 flavors of sea quarks yielded confusing results.
In a few years, the low moments of quark densities from 2+1- and 2+1+1-flavor simulations should become good
enough to incorporate into the traditional fits of experimental data.
For collider phenomenology, however, the real challenge for lattice QCD is to compute similar moments of the
gluon density, which are less well constrained by low-energy experiments.

\section{QCD Thermodynamics}
\label{sec:thermo}

The previous sections consider isolated hadrons at zero temperature.
Soon after the Big Bang, however, the universe was much hotter than it is now.
In neutron stars, for example, the baryon density is much higher than in normal nuclear matter.
These phenomena have motivated the study of the thermodynamics of QCD.
Even within lattice gauge theory, thermodynamics is a vast subject~\cite{DeTar:2009ef,Fodor:2009ax}, so this
review touches only on some of the more fascinating aspects.

Thermodynamics starts with thermal averages in the canonical ensemble
\begin{equation}
    \langle\bullet\rangle = \frac{\Tr\left[\bullet\,e^{-\hat{H}/T}\right]}{\Tr e^{-\hat{H}/T}},
    \label{eq:thermal}
\end{equation}
where $T$ is the temperature, and the traces $\Tr$ are over the Hilbert space of the QCD Hamiltonian
$\hat{H}$.
In fact, the average on the left-hand side of Equation~\ref{eq:thermal} is precisely that of
Equation~\ref{eq:Z}; the time extent $N_4$ specifies the temperature $T=(N_4a)^{-1}$.
The eigenstates of $\hat H$, which are single hadrons and multiparticle states composed of hadrons, do not
change with $T$, but as $T$ increases, the vacuum no longer dominates the way it does in
Equations~\ref{eq:2pt-m}--\ref{eq:3pt}, and multihadron states begin to play a role in the thermal average.
\phantom{\cite{Borsanyi:2010bp}}
\begin{figure}
    \vspace*{-4pt}
    \centering
    \includegraphics[width=0.8\textwidth]{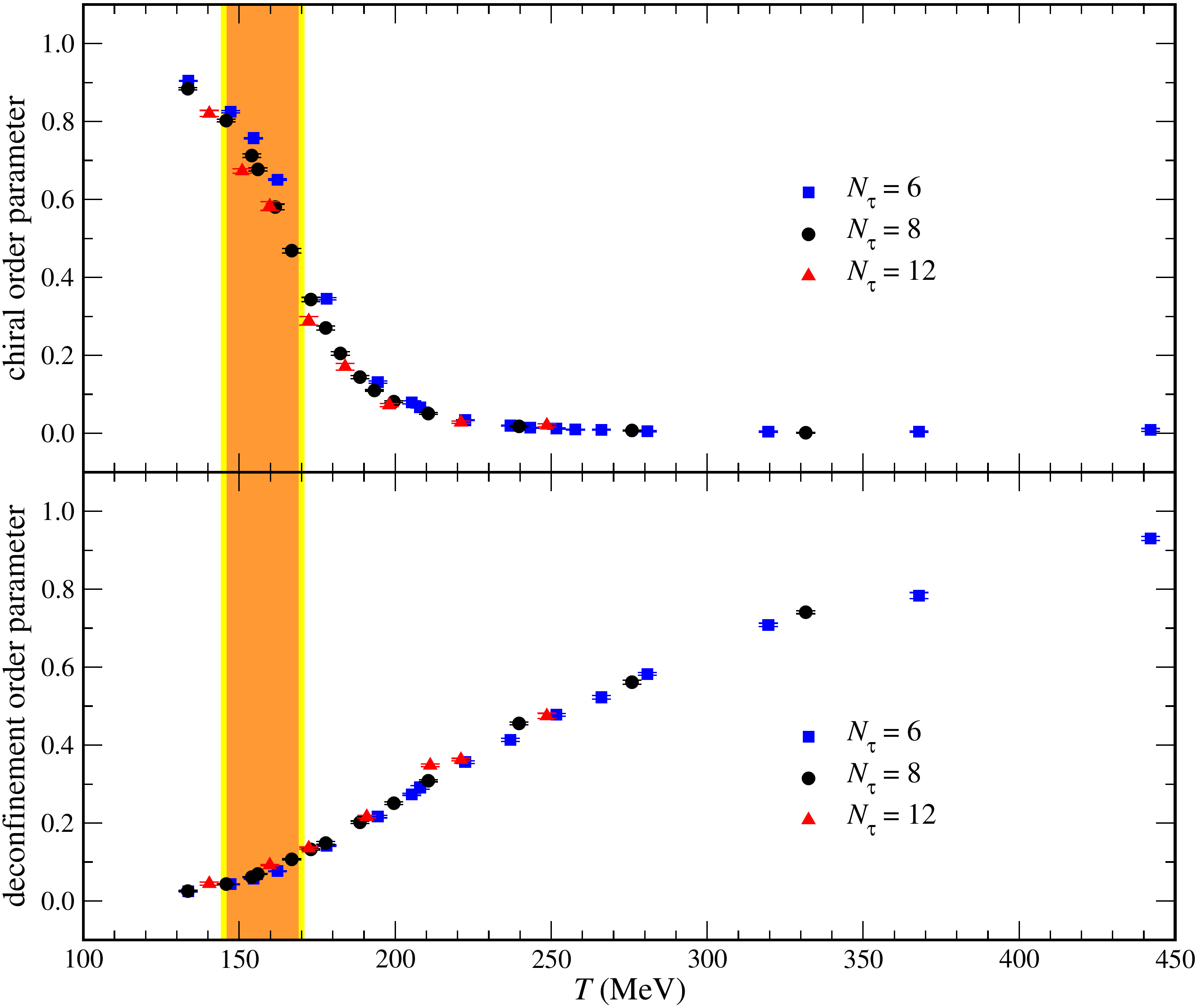}
    \vspace*{-7pt}
	\caption[fig:Tc]{Order parameters for deconfinement (\emph{bottom}) and chiral symmetry restoration 
        (\emph{top}), as a function of temperature. 
        The physical temperature $T=(N_\tau a)^{-1}$, where $a$ is the lattice spacing and $N_\tau=N_4$.
        Agreement for several values of $N_\tau$ thus indicates that discretization effects from the lattice 
        are under control.
        Data are from Reference~\citeonline{Bazavov:2011nk}.}
	\label{fig:Tc}
\end{figure}

The simplest observables are quantities suach energy, pressure, entropy density, and order parameters
sensitive to symmetry breaking.
The thermal state can either restore a spontaneously broken symmetry of the vacuum or be a state of broken
symmetry itself.
Of course, the (approximate) symmetry of the Hamiltonian remains intact.
Figure~\ref{fig:Tc} shows order parameters for deconfinement and chiral symmetry restoration, as the
temperature increases from normal hadronic matter to a phase known as the quark-gluon plasma.
Both order parameters change dramatically for a temperature around
145--170~MeV~\cite{Borsanyi:2010bp,Bazavov:2011nk}, but neither, especially deconfinement, exhibits the sharp
change characteristic of a phase transition.
Studying a whole suite of thermodynamic observables confirms that the transition is a smooth
crossover~\cite{Aoki:2006we,Bazavov:2009zn}.
This result came as a surprise, and the next two paragraphs explain why.

The crossover means that as the early universe cooled, hot matter gradually became more and more like a gas
of distinct hadrons.
With a first-order phase transition, on the other hand, bubbles of the hadronic phase would have formed
inside the quark-gluon plasma.
Without a real phase transition, the quark-gluon plasma is not necessarily a fluid of quasi-free quarks and
gluons.
The eigenstates in Equation~\ref{eq:thermal} remain color singlets, but a thermal medium can be 
qualitatively different. 
First, thermal fluctuations encompass states with many overlapping hadrons, so color can propagate from one
hadron to the next, as if deconfined.
Second, the thermal average applies nearly equal Boltzmann weights to states of both parities, so chiral
symmetry can be restored in the thermal average, even though the vacuum breaks~it.

\begin{figure}
    (\emph{a})~\includegraphics[width=0.40\textwidth]{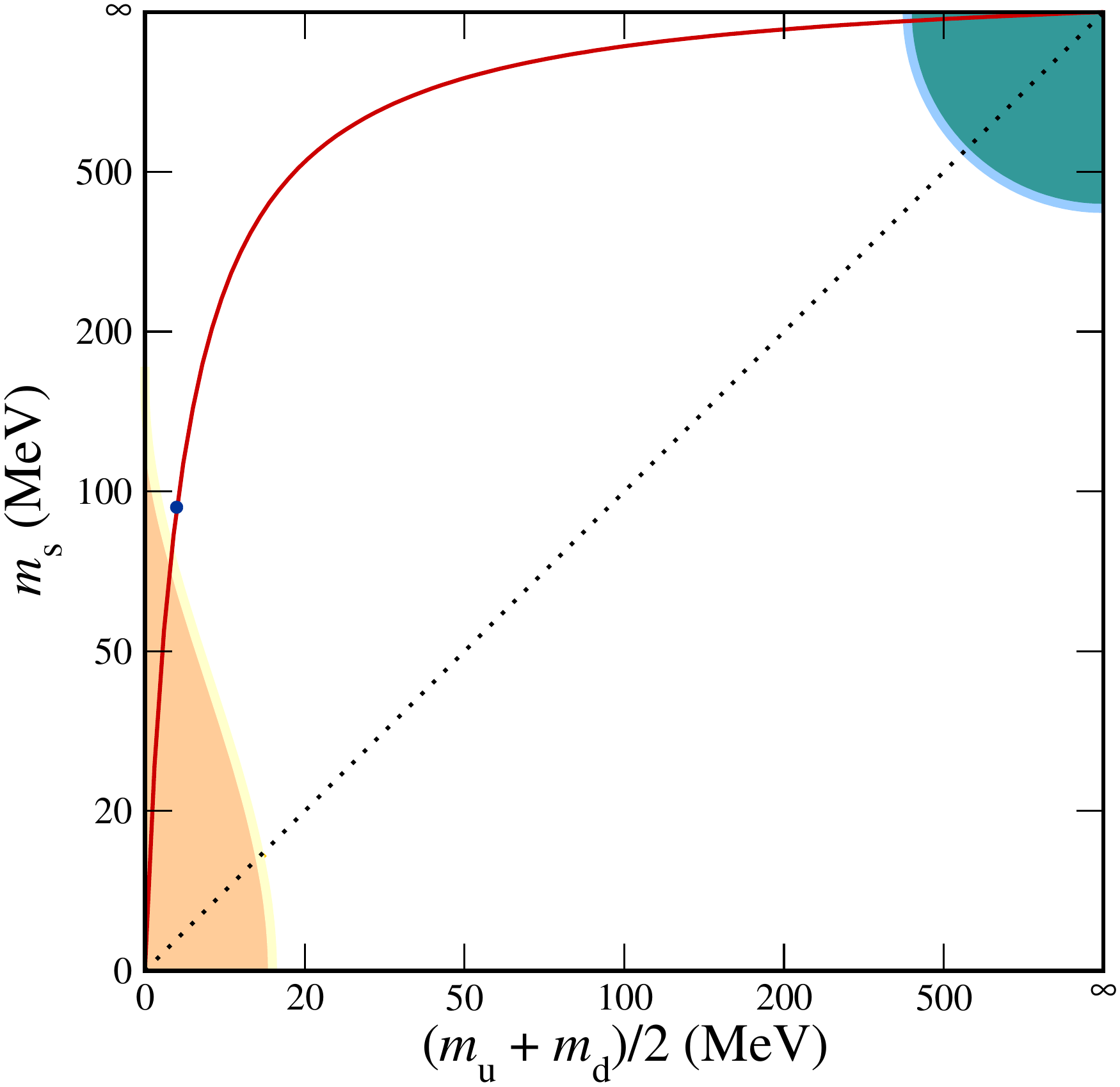}
    \hfill
    (\emph{b})~\includegraphics[width=0.40\textwidth]{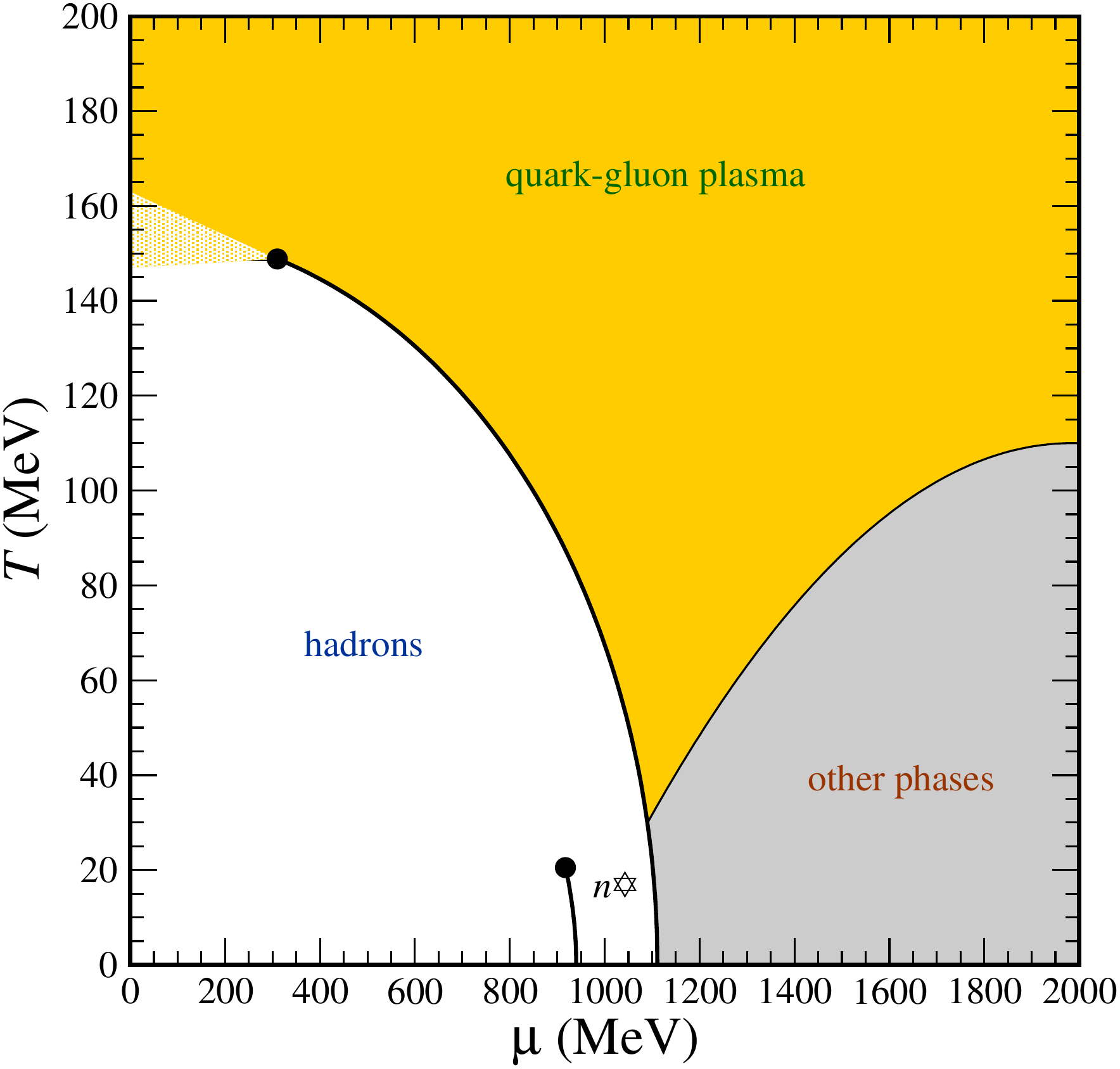}
    \caption[fig:phase]{QCD phase diagrams.
        (\emph{a})~The $\mstr$-$\frac{1}{2}(\mup+\mdn)$ plane at $(\mu,T)=(0,T_c)$, showing the order of the 
        transition.
        The shaded regions at very small and nearly infinite masses are first order; the red line shows the 
        physical ratio of $2\mstr/(\mup+\mdn)$.
        (\emph{b})~The $\mu$-$T$ plane, showing the crossover at small $\mu$ determined from lattice QCD.
        The neutron star (denoted $n\star$) and other phases are expected, but lattice QCD is not yet in a 
        position to provide useful information.}
    \label{fig:phase}
\end{figure}

The nature of the QCD phase transition is influenced by the physical values of the up, down, and strange
quark masses.
For vanishing quark masses, the transition would be first order, but as the masses are increased, the
strength of the transition diminishes.
As depicted in Figure~\ref{fig:phase}\emph{a}, the physical quark masses (Table~\ref{tab:q}) are just large
enough to render the transition a crossover.
If the light quark masses---crucially $\mstr$---were around half their physical size, the universe would 
have cooled through a first-order transition.
Before lattice QCD established these results, the conventional wisdom was that the quark masses are somewhat
larger than shown in Table~\ref{tab:q}, yet small enough to remain in the first-order basin of massless
quarks.

At nonzero baryon density (chemical potential~$\mu\neq0$), the fermion determinant becomes complex, which is
an obstacle to importance sampling.
This restricts lattice QCD calculations to small $\mu$.
It is thought that the transition becomes first order for $\mu\sim$ few~100~MeV
(Figure~\ref{fig:phase}\emph{b}) \cite{Levkova:2012jd}, but the matter is not yet
settled~\cite{deForcrand:2008vr}.

\section{Summary and Outlook}
\label{sec:sum}

The topics discussed above demonstrate that we have learned a great deal in this century about QCD from 
lattice gauge theory.
The twenty-first century is still young, and the prospects for learning more are bright.

Although the mass spectrum of the lowest-lying hadrons has been well verified, it will be interesting to 
extend the calculations to excited states~\cite{Bulava:2009jb,Edwards:2011jj} and even to small nuclei such 
as the deuteron~\cite{Beane:2011iw} or the $H$ dibaryon~\cite{Beane:2010hg,Inoue:2010es}.
Beyond QCD, one may wonder whether nature uses gauge theories to generate quark, lepton, and weak boson 
masses~\cite{DelDebbio:2010zz}.

Most of the calculations related to flavor physics are entering an industrial phase, where the objective is
higher and higher precision.
An exception is the measurement of direct \CP\ violation in the kaon system.
This calculation requires a two-pion final state, and although the formalism for handling this state has long
been available~\cite{Lellouch:2000pv}, only now have $K\to\pi\pi$ amplitudes become
feasible~\cite{Blum:2011pu,Blum:2011ng}.
These finite-volume techniques are related to methods~\cite{Luscher:1990ux} for computing scattering
lengths~\cite{Beane:2006gf,Torok:2009dg}, which have many applications in hadronic physics.

\appendix

\section*{Appendix: Tools}
\label{sec:tools}
\addcontentsline{toc}{section}{Appendix: Tools}

Research in lattice QCD requires computer time and software.
Through several efforts around the world, these needs pose lower obstacles than ever before.
Several groups have made documented software available so that new programs can be modeled after existing
code, rather than being built from scratch.
Furthermore, many groups make ensembles of lattice gauge fields available, principally via the International
Lattice Data Grid (ILDG) (\url{http://www.usqcd.org/ildg/}).
In exchange for a suitable citation of an article describing the content of the ensembles, anyone can use
these simulation data for his or her own physics analyses.
In many cases, even more ensembles are available from collaborations with newer ensembles under generous
terms: Most of these collaborations have some core physics analyses but are happy if the expensive simulation
data are mined for more results.

The ILDG has portals in 
Australia (\url{http://cssm.sasr.edu.au/ildg/}),
Japan (\url{http://ws.jldg.org/QCDArchive/}),
continental Europe (\url{http://hpc.desy.de/ldg/}),
the United Kingdom (\url{http://www.gridpp.ac.uk/qcdgrid/}),
and the United States (\url{http://www.usqcd.org/ildg/}).
The technical underpinnings are described in Reference~\citeonline{Beckett:2009cb}.
Further ensembles are available from the Gauge Connection~(\url{http://qcd.nersc.gov/}) and
the QCDOC Gauge Field Configuration Archive~(\url{http://lattices.qcdoc.bnl.gov/}).

Publicly available software can be obtained from the USQCD Collaboration 
(\url{http://www.usqcd.org/usqcd-software/}).
Newcomers should start with one of the applications packages:
\textsc{chroma}, \textsc{cps}, \textsc{milc}, or \textsc{FermiQCD}.
A~useful tutorial on this software has been put together by Jo\'o~(\citeonline{Joo:INT07}; 
contains slides only).

Two kinds of computing are important to lattice gauge theory, \emph{capability} and \emph{capacity}.
One needs access to computers with the greatest capability---those able to run large-memory jobs with huge
appetite in CPU time---to generate the ensembles of lattice gauge fields.
On these ensembles an analysis consists of a huge number of small-to-medium computing demands; this step
requires computers with high capacity.
Many university groups have access to computers of sufficiently high capacity to analyze the publicly
available ensembles.

\section*{Acknowledgments}

The author thanks Jimmy Juge, Julius Kuti, \& Colin Morningstar for supplying the data plotted in 
Figure~\ref{fig:V}, 
Eric Gregory \& Craig McNeile for providing numerical values for their $\eta$ and $\eta'$ masses 
(Figure~\ref{fig:spectrum}), and 
Frithjof Karsch for help with Figure~\ref{fig:Tc}.
Fermilab is operated by Fermi Research Alliance, LLC, under contract number DE-AC02-07CH11359 with
the U.S. Department of Energy.


\addcontentsline{toc}{section}{Literature Cited}

\end{document}